\documentclass{iaus}
\usepackage{graphicx}
\usepackage{units}

\title[~~A MOCASSIN Model of NGC 4449] 
{A Multi-wavelength MOCASSIN Model of the Magellanic-type Galaxy NGC~4449}

\author[O. \L. Karczewski \etal]   
{O. \L. Karczewski$^1$, M. J. Barlow$^2$, M. J. Page$^3$ \and S. C. Madden$^4$ }

\affiliation{
$^1$Department of Physics and Astronomy, University College London, \\
Gower Street, London, WC1E 6BT, UK \\
e-mail: {\tt olk@star.ucl.ac.uk} \\[\affilskip]
$^2$Department of Physics and Astronomy, University College London, \\
Gower Street, London, WC1E 6BT, UK \\
e-mail: {\tt mjb@star.ucl.ac.uk} \\[\affilskip]
$^3$Mullard Space Science Laboratory, University College London, \\
Holmbury St. Mary, Dorking, RH5 6NT, UK \\
e-mail: {\tt mjp@mssl.ucl.ac.uk} \\[\affilskip]
$^4$CEA Saclay Service d'Astrophysique, L'Orme des Merisiers Bat 709, \\
91191 Gif-sur-Yvette Cedex, France \\
e-mail: {\tt suzanne.madden@cea.fr}}

\pubyear{2011} 
\volume{284}  
\pagerange{1--12}
\jname{The Spectral Energy Distribution of Galaxies}
\editors{R.J. Tuffs \&  C.C.Popescu, eds.}
\begin{document}

\maketitle

\begin{abstract}
We use the photoionisation and dust radiative transfer code MOCASSIN to create a model of the dwarf irregular galaxy NGC 4449. The best-matching model reproduces the global optical emission line fluxes and the observed spectral energy distribution (SED) spanning wavelengths from the UV to sub-mm, and requires the bolometric luminosity of 6.25$\,\times\,$10$^9$ $L_\odot$ for the underlying stellar component, $M_{\mathrm{dust}}$/$M_{\mathrm{gas}}$ of 1/680 and $M_{\mathrm{dust}}$ of 2.2$\,\times\,$10$^6$ $M_\odot$.

\keywords{dust, extinction, galaxies: dwarf, galaxies: general, galaxies: individual (NGC 4449), galaxies: stellar content, methods: numerical}
\end{abstract}

\firstsection 
\section{Introduction}

The spectral energy distribution (SED) of a galaxy emerges as a combination of
the underlying stellar emission, the degree of reprocessing taking
place in the interstellar medium (ISM) and thermal or non-thermal emission due to
dynamical or evolutionary processes. As a consequence, galaxies emit across the entire
range of the electromagnetic spectrum and the distribution of the energy output as a function
of wavelength provide clues to the galaxy's past and present. Numerical models that
reproduce the observed global properties of a galaxy over a wide range of
wavelengths can be useful in understanding the physical and chemical nature
of the underlying stellar and dust components. Here we present a MOCASSIN model of the
dwarf irregular galaxy NGC~4449.

MOCASSIN \cite[(Ercolano \etal\ 2003, 2005, 2008)]{Ercolano2003,Ercolano2005,Ercolano2008} 
is a fully 3-dimensional photoionisation and dust radiative transfer code.
Recent applications of the code include modelling the planetary nebula
NGC~6302 \cite[(Wright \etal\ 2011)]{Wright2011} and studying dust formation
by SN~2008S \cite[(Wesson \etal\ 2010)]{Wesson2010}.
Modelling entire galaxies is technically challenging due to their large physical
sizes and the high level of complexity of the galactic environments. Dwarf galaxies are ideal
candidates due to their intrinsically small sizes, relatively simple star formation
histories and the low degree of processing of their interstellar material.

\begin{figure}
\vspace{2mm}
\begin{center}
 \includegraphics[width=5.4cm]{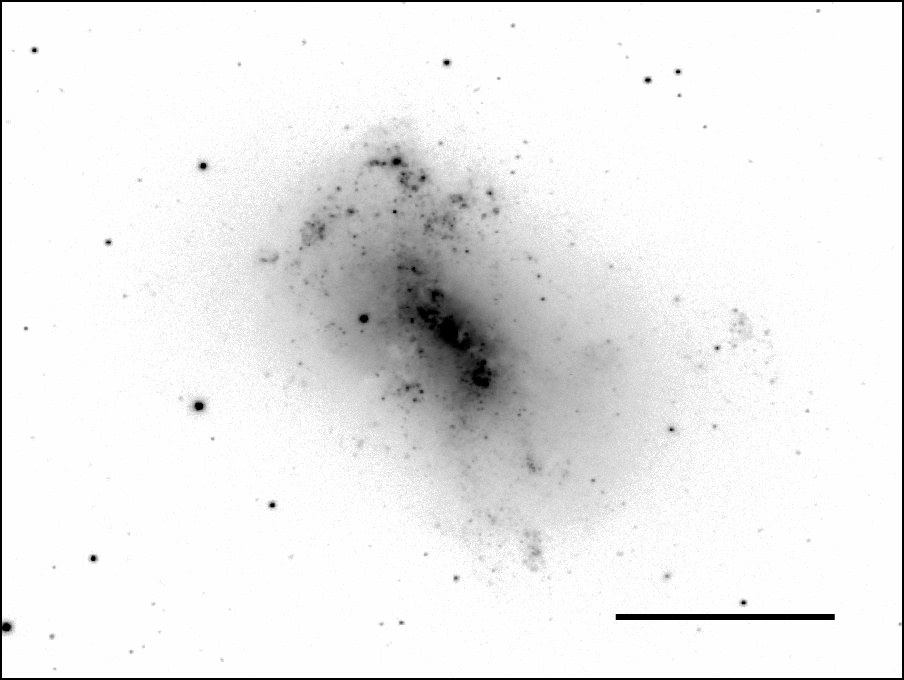}
 \caption{SDSS \textit{r}-band image of NGC~4449. North is up, east is to the left. The image is centred at 12$^{\rm h}$28$^{\rm m}$11$.\!\!^{\rm s}$1, +44$^{\circ}$05$^{\prime}$37$^{\prime\prime}$ (J2000). The bar is 2$^{\prime}$ in length.}
   \label{fig1}
\end{center}
\end{figure}

NGC~4449 (Fig.\,\ref{fig1}) is a metal-poor magellanic-type irregular galaxy. It is not
a typical dwarf irregular: at the distance of 4.21 Mpc its H~{\sc i} envelope extends to a
radius of 14.2 kpc (11$.\!\!^{\prime}$6), equal to approximately 6$R_{25}$
\cite[(Swaters \etal\ 2002)]{Swaters2002},
and exhibits a counter-rotating core. It was selected for this study
for its extensive coverage of photometric measurements for
wavelengths ranging from the UV to sub-mm.
Recently, new FIR observations were acquired by the {\it Herschel Space Observatory}
\cite[(Pilbratt \etal\ 2010)]{Herschel} as part of the 
Guaranteed Time Key Project: Dwarf Galaxy Survey (PI: Suzanne Madden).

The emission line fluxes, used to constrain the models, were derived from global spectra,
acquired with the drift scanning technique, and are representative of the
entire galaxy \cite[(Kobulnicky \etal\ 1999)]{Kobulnicky1999}. 

\vspace{-2mm}

\section{Methods and Results}

\begin{figure}[b]
\vspace{2mm}
\begin{center}
 \includegraphics[width=8.8cm]{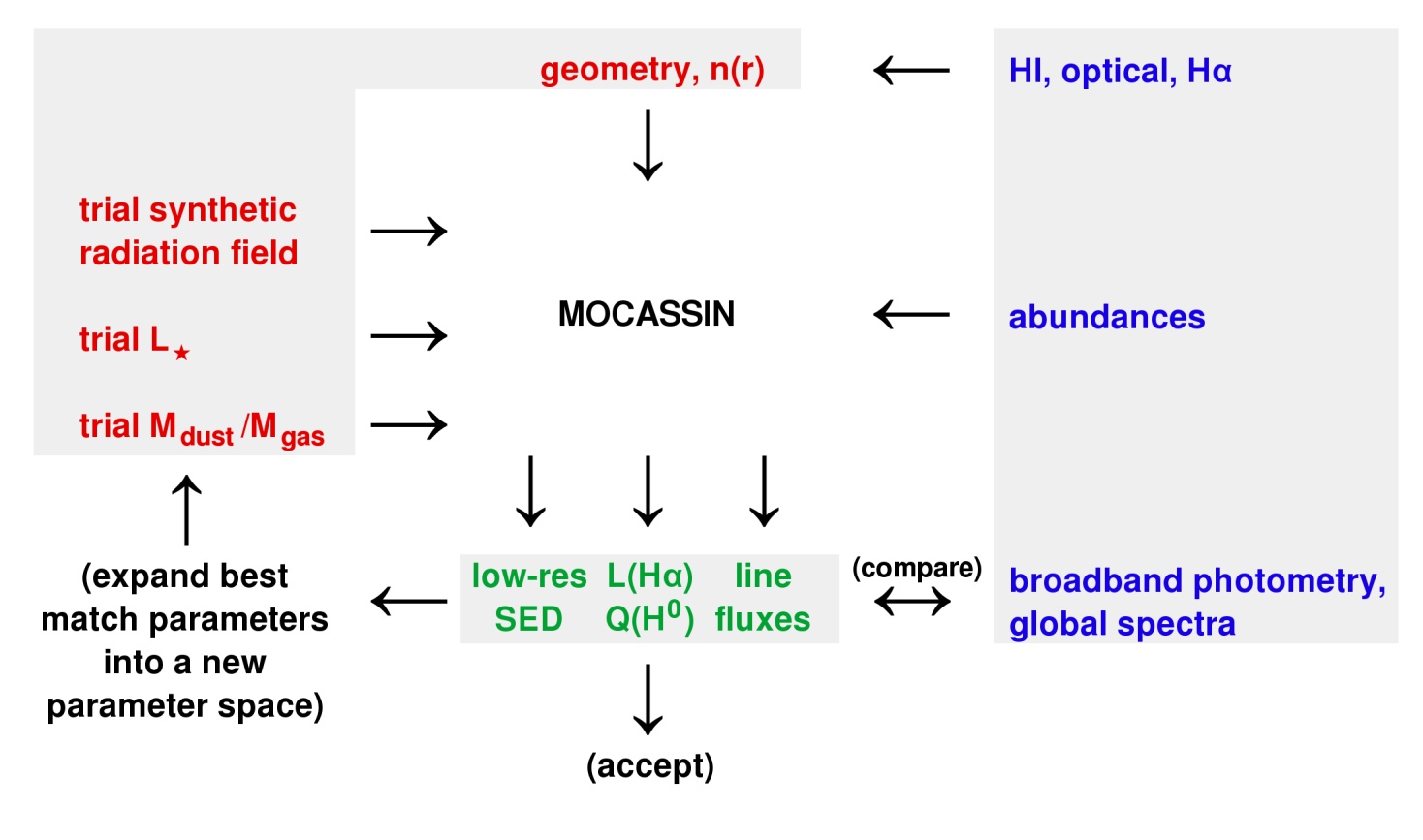} \tabularnewline{}
 \vspace{-0.2 cm}
 \caption{A schematic summary of the modelling method used.
The shaded regions group together the inputs to
MOCASSIN (left) the outputs (middle) and the observational constrains (right).}
   \label{fig2}
\end{center}
\end{figure}

Our MOCASSIN simulations use 0.5 million grid cells to establish the observed H~{\sc i} profile of
\cite[Swaters \etal\ (2002)]{Swaters2002} within the inner 3.3 kpc (2$.\!\!^{\prime}$7), enclosing
the visible part of NGC~4449 and about a half of its total H~{\sc i} mass. The geometry,
gas density distribution and the observed abundances are fixed inputs. Trial families of simulations 
are run and a range of $M_{\mathrm{dust}}$/$M_{\mathrm{gas}}$, bolometric stellar luminosities $L_\star$
and input stellar radiation fields are tested. A parameter set providing the best match to the
shape of the low-resolution SED, the global emission line fluxes and the integrated {H}$\alpha$ luminosity
\cite[(Hunter \etal\ 1999)]{Hunter1999} is expanded into a new, higher resolution
parameter space and the procedure is iterated until a satisfactory
match with all observables is obtained. Fig.~\ref{fig2} gives
a schematic summary of this method.

\begin{figure}[h]
\begin{center}
 \includegraphics[scale=0.92]{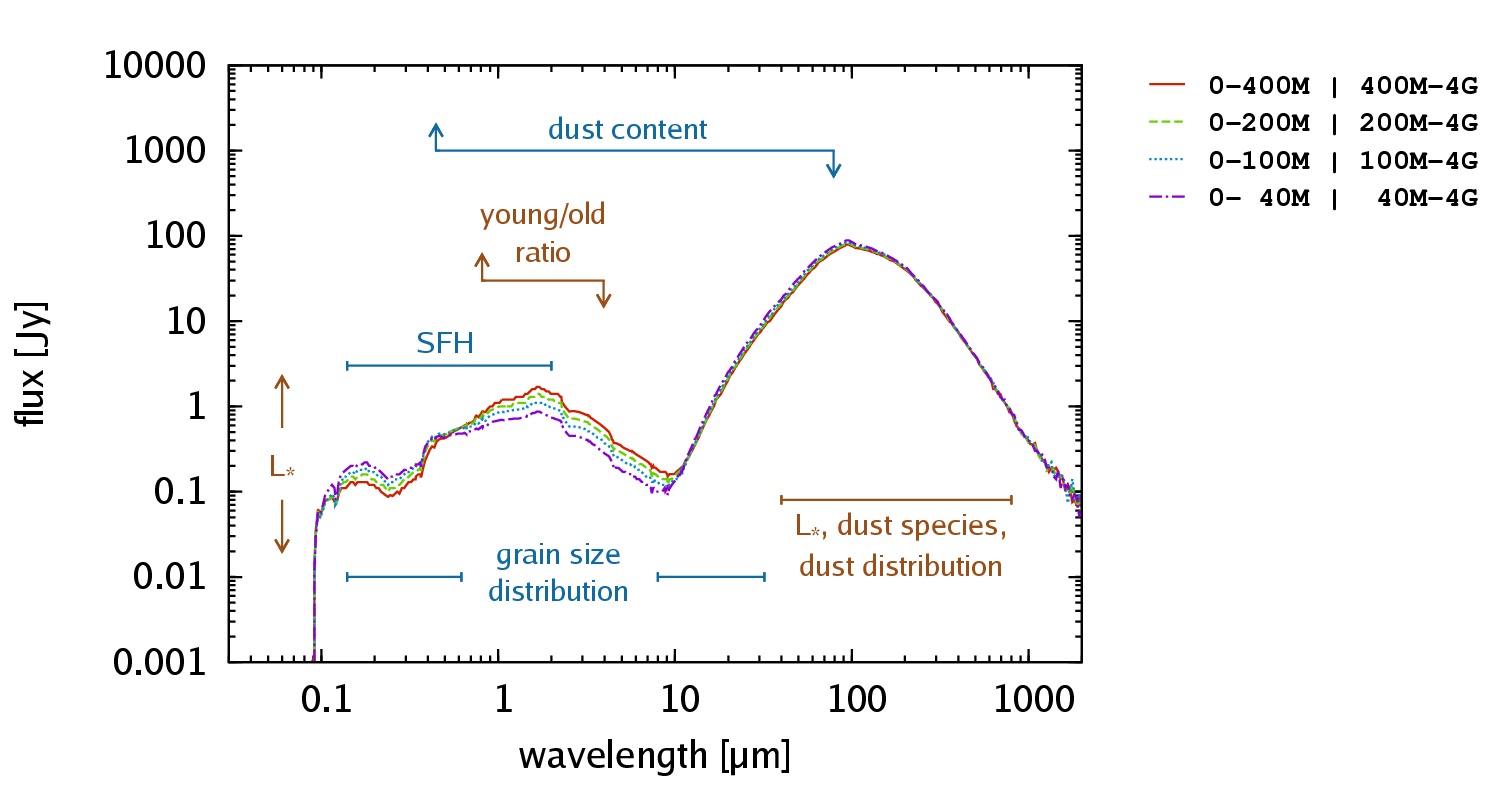} 
\vspace{-1.5mm}
 \caption{A family of predicted SEDs for four trial star formation scenarios
assuming two episodes of continuous star formation with two different star
formation rates. All input parameters are identical except for the length
of the two star formation episodes (see legend for details).
The approximate wavelength ranges affected by the various free parameters are indicated.}
   \label{fig3}
\end{center}
\end{figure}

The stellar radiation field is the most crucial input, as it results from the assumed
star formation history. In this work, two continuous episodes of constant star formation activity
were assumed and an initial (and subsequently expanded, according to Fig.~\ref{fig2}) set of 45 templates generated with STARBURST99 \cite[(Leitherer \etal\ 1999; version 2011-01-17)]{Leitherer1999} was tested. A small subset of the resulting SEDs is shown in Fig.~\ref{fig3}.

\begin{figure}[h]
\begin{center}
 \includegraphics[scale=0.9]{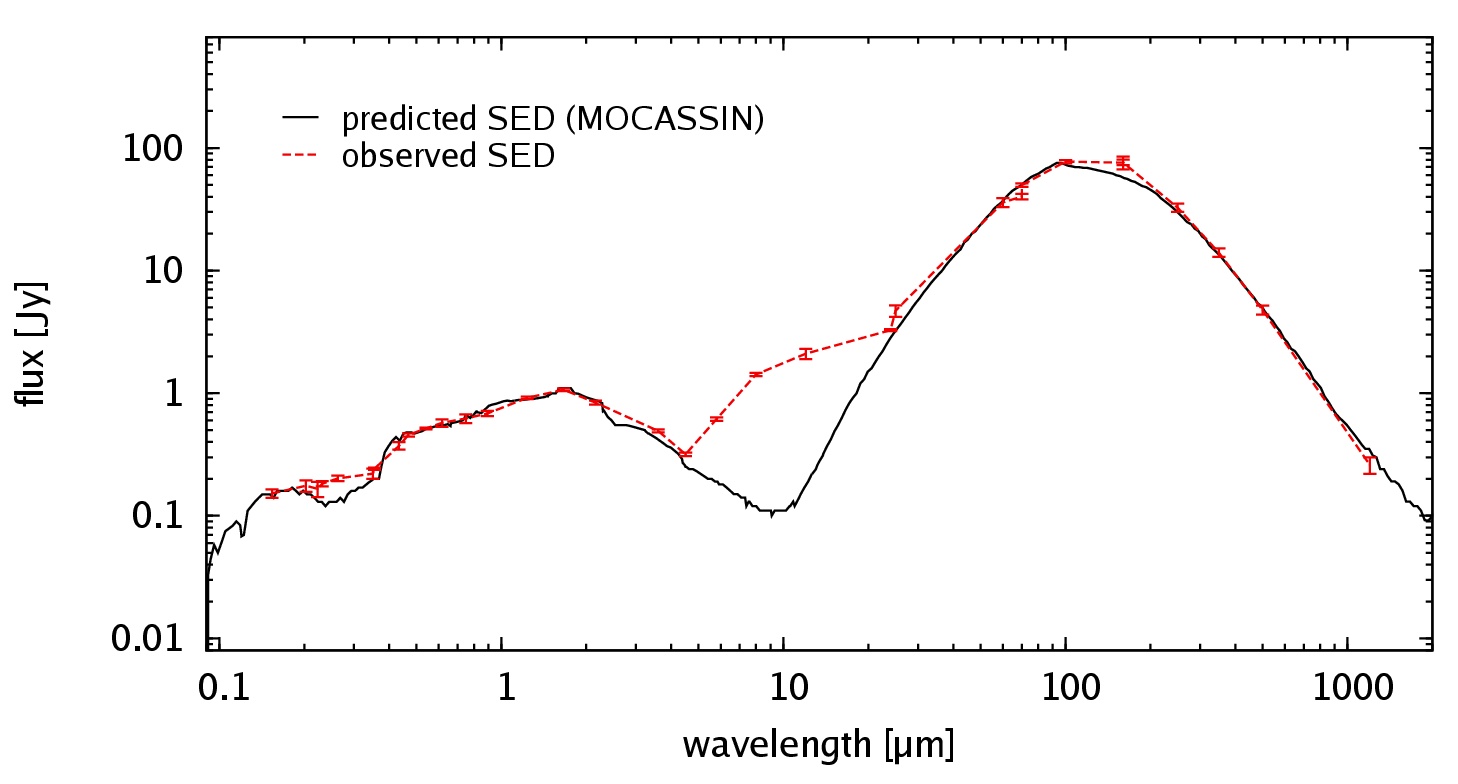} 
\vspace{-1mm}
 \caption{The predicted SED of the best-matching MOCASSIN model of NGC~4449. PAHs are not included in this model.
The observed SED is based on the archive data ({\it GALEX} in the UV, {\it Swift} in the UV/optical, SDSS in the optical, 2MASS in the NIR) and the published photometry (\cite[Engelbracht \etal\ 2008]{Engelbracht2008}, \cite[Hunter \etal\ 1986]{Hunter1986}, \cite[B\"{o}ttner \etal\ 2003]{Bottner2003}; at longer wavelengths). In the FIR new {\it Herschel} PACS and SPIRE measurements are included (Remy \etal, in prep.).
}
   \label{fig4}
\end{center}
\end{figure}

To break the degeneracy it is helpful to ask if both \textit{(i)} the shape of the observed SED and
 \textit{(ii)} the total emergent $L(\textrm{H}\alpha)$ for a particular star formation
scenario \textit{can} be matched by varying $M_{\mathrm{dust}}$/$M_{\mathrm{gas}}$
and/or $L_\star$, whose effect on the resulting SED is relatively simple to predict.
This condition allows for more than 90 per~cent of the initial
templates to be rejected at an early stage. The final best-matching model is selected by
inspecting the predicted emission line fluxes. For NGC~4449 the best-matching SED is presented
in Fig.~\ref{fig4} and the derived physical quantities are given in Table~\ref{tab1}.

\section{Discussion}

The gas and dust masses derived from our MOCASSIN model of NGC~4449 are consistent with earlier estimates (\cite[Hunter \etal\ 1999]{Hunter1999}; \cite[Engelbracht \etal\ 2008]{Engelbracht2008}). Our best-matching star formation scenario broadly agrees with the presence of an old (3--5 Gyr) and a young population of stars, and continuous star formation over the last 1~Gyr (\cite[Bothun 1986]{Bothun1986}; \cite[Martin \& Kennicutt 1997]{Martin1997}; \cite[Annibali \etal\ 2008]{Annibali2008}). However, the predicted present-day star formation rate is only one-fourth of the estimate based on
the H$\alpha$ luminosity \cite[(Hunter \etal\ 1999)]{Hunter1999}. The discrepancy between these two
independent estimates highlights the differences in the adopted initial mass function (IMF)
as well as the limitations of the two-episode model.

The presented technique assumes spherical symmetry for NGC~4449 and a two-episode star formation history, both of which are
great simplifications. Also, at present PAHs are not included in the models and, as a consequence, the 4--20~$\mu$m range is not correctly predicted. Nevertheless, most of the global parameters are predicted to within the observational uncertainties, and therefore this technique may prove useful in studying the characteristics of dust and the underlying stellar components in other dwarf galaxies.

\begin{table}
\vspace{0.2cm}
\begin{center}
\begin{tabular}{l@{\hspace{8mm}}cc}
\cline{1-3}\noalign{\medskip}

parameter & \multicolumn{2}{c}{MOCASSIN model of NGC~4449}\tabularnewline \hline\noalign{\smallskip}
$L_\star$ & \multicolumn{2}{c}{6.25$\,\pm\,$0.25$\,\times\,$10$^9$ $L_\odot$} \tabularnewline
$M_\star$ & \multicolumn{2}{c}{1.2$\,\pm\,$0.1$\,\times\,$10$^9$ $M_\odot$} \tabularnewline
$M_\mathrm{gas}$ & \multicolumn{2}{c}{1.5$\,\pm\,$0.2$\,\times\,$10$^9$ $M_\odot$} \tabularnewline
$M_\mathrm{dust}$ & \multicolumn{2}{c}{2.2$\,\pm\,$0.2$\,\times\,$10$^6$ $M_\odot$} \tabularnewline
$M_{\mathrm{dust}}$/$M_{\mathrm{gas}}$ & \multicolumn{2}{c}{ 1/680 (1/850 to 1/540)} \tabularnewline
dust composition & \multicolumn{2}{c}{100\% amorphous carbon} \tabularnewline
dust grain sizes & \multicolumn{2}{c}{0.005--1~$\mu$m, $n \propto a^{-3.5}$} \tabularnewline
\noalign{\medskip}
SF episodes & 6$\,\pm\,$2~Gyr to 120$\,\pm\,$40~Myr & 120$\,\pm\,$40~Myr  to present \tabularnewline
relative $M_\star$ & 120 & 1 \tabularnewline
$<$SFR$>$ & 0.2 to 0.3 $M_\odot$\,yr$^{-1}$ & 0.06 to 0.12 $M_\odot$\,yr$^{-1}$ \tabularnewline
\noalign{\smallskip}\cline{1-3}
\end{tabular}
\vspace{0.2cm}
  \caption{Summary of global parameters derived from the best-matching MOCASSIN model of NGC~4449 assuming two continuous episodes of star formation. The global emission line fluxes are matched to within 20 per~cent or better, with the exception of the sulphur lines [S~{\sc ii}] $\lambda\lambda$6717,31, which are overpredicted by a factor of four.}
  \label{tab1}
\end{center}
\end{table}

\end{document}